\documentclass[12pt]{article}
\usepackage[latin1]{inputenc}
\usepackage[british]{babel}
\usepackage{cmap}
\usepackage{lmodern}

\usepackage{amssymb, amsmath, amsthm}
\usepackage[a4paper,top=25mm,bottom=25mm,left=25mm,right=25mm]{geometry}
\usepackage{etex}
\usepackage{ragged2e}

\usepackage{authblk} % for headings
\usepackage{pifont}
\usepackage{graphicx}
\usepackage[usenames,dvipsnames,svgnames,table]{xcolor}
\usepackage[figuresright]{rotating}
\usepackage{xtab} % tackle the long tables
\usepackage{longtable} % tackle the long tables
\usepackage{multirow}
\usepackage{footnote}
\usepackage[stable]{footmisc}
\usepackage{chngpage} % allows for temporary adjustment of side margins
\usepackage{pdflscape} % landscape environment
\usepackage[nottoc,notlot,notlof]{tocbibind} % includes everything in ToC

\usepackage{pgfplots}
\pgfplotsset{compat=1.14}
\usepackage{setspace}
\usetikzlibrary{backgrounds}
\usepackage{xcolor}
\makesavenoteenv{tabular}
\usepackage{tabularx}
\usepackage{booktabs}
\usepackage{threeparttable}
\usepackage[referable]{threeparttablex} % footnotes in tabu
\newcolumntype{R}{>{\raggedleft\arraybackslash}X}
\newcolumntype{L}{>{\raggedright\arraybackslash}X}
\newcolumntype{C}{>{\centering\arraybackslash}X}
\newcolumntype{M}[1]{>{\centering\arraybackslash}m{#1}}
\newcolumntype{A}{>{\columncolor{gray!25}}C}
\newcolumntype{a}{>{\columncolor{gray!25}}c}
\usepackage{wasysym}
\usepackage{array}
\newcolumntype{K}[1]{>{\centering\arraybackslash$}p{#1}<{$}}

\newlength{\tablen}

\usepackage{dcolumn} % alignment to decimal points
\newcolumntype{.}{D{.}{.}{-1}}

\usepackage{tikz}
\usetikzlibrary{arrows, calc, matrix, patterns, positioning, trees, decorations.markings, automata}
\usepackage[semicolon]{natbib}
\usepackage[hyphens]{url}
\usepackage{hyperref} % [hidelinks]
\hypersetup{
  colorlinks   = true,    % Colours links instead of ugly boxes
  urlcolor     = blue,    % Colour for external hyperlinks
  linkcolor    = blue,    % Colour of internal links
  citecolor    = red      % Colour of citations
}
\usepackage{microtype}
\usepackage[justification=centerfirst]{caption} % [justification=centering]

\usepackage[labelformat=simple]{subcaption}

\DeclareCaptionLabelFormat{parenthesis}{(#2)}
\makeatletter
\renewcommand\p@subfigure{\arabic{figure}.}
\makeatother

\DeclareCaptionLabelFormat{parenthesis}{(#2)}
\makeatletter
\renewcommand\p@subtable{\arabic{table}.}
\makeatother

\usepackage{enumitem}
\setlist[itemize]{leftmargin=2.5\parindent}
\setlist[enumerate]{leftmargin=2.5\parindent}

\theoremstyle{plain}
\theoremstyle{definition}

\newtheorem{definition}{Definition}[section]
\newtheorem{example}{Example}[section]

\theoremstyle{remark}

\newcommand{\down}{\textcolor{BrickRed}{\rotatebox[origin=c]{270}{\ding{212}}}}
\newcommand{\up}{\textcolor{PineGreen}{\rotatebox[origin=c]{90}{\ding{212}}}}
\newcommand{\const}{\textcolor{darkgray}{\footnotesize{\CIRCLE}}}

\def\keywords{\vspace{.5em} % Add keywords
{\noindent \textit{Keywords}:\,}}

\def\JEL{\vspace{.5em} % Add keywords
{\noindent \textbf{\emph{JEL} classification number}:\,}}

\def\AMS{\vspace{.5em} % Add keywords
{\noindent \textbf{\emph{MSC} class}:\,}}

\author{\href{https://scholar.google.hu/citations?user=b4So0I0AAAAJ&hl=hu}{D\'ora Gr\'eta Petr\'oczy}\thanks{~E-mail: \emph{doragreta.petroczy@uni-corvinus.hu}}} %Address: H-1093 Budapest, Fõvám Square 8. }
\affil{Corvinus University of Budapest (BCE) \\
Department of Finance}
\affil{Budapest, Hungary}
\title{An alternative quality of life ranking \\ on the basis of remittances
}
\date{\today}

\begin{document}

\maketitle

\begin{abstract}
\noindent
Remittances provide an essential connection between people working abroad and their home countries. This paper considers these transfers as a measure of preferences revealed by the workers, underlying a ranking of countries around the world. In particular, we use the World Bank bilateral remittances data of international salaries and interpersonal transfers between 2010 and 2015 to compare European countries. The suggested least squares method has favourable axiomatic properties. Our ranking reveals a crucial aspect of quality of life and may become an alternative to various composite indices.

\JEL{C44, F22, O57}
% Operations Research, Statistical Decision Theory
% International Migration
% Comparative Studies of Countries 

\AMS{15A06, 62F07, 91B82}
% Linear equations
% Ranking and selection
% Statistical methods; economic indices and measures

\keywords{Paired comparison; remittances; international migration; quality of life; country ranking}
\end{abstract}
%\clearpage

\section{Introduction} \label{Sec1}

Researchers have maintained an interest in measuring the quality of life in various geographic areas, since the 1930s, when President Hoover's Committee on Social Trends issued its report  ``Recent Social Trends in the US'' \citep{Wish1986}. Country rankings seem to be increasingly popular in economics and can often have a considerable impact on politicians and government strategies. In the recent decades, scientists have recommended several alternative approaches to define and measure the quality of life, see \citep{DienerSuh1997} for a summary. However, most of them are composite indices, a construction that remains highly controversial due to the arbitrary selection of criteria and ad hoc choice of component weights \citep{Ravallion2012a}. 
 While robustness check may provide a kind of remedy \citep{FosterMcGillivraySeth2013}, there exists an alternative solution, that is, to apply a parameter-free algorithm on an appropriate dataset.

It is widely argued that people and their capabilities should be the ultimate criteria for assessing a country's development, not economic indicators alone \citep{Sen1985,Sen1992}. One way to measure the perceptions of people is to observe their decisions on important questions of life such as working abroad. Although migration emanates from the desire to improve livelihood, it requires a certain development level, so it is not the poorest who migrate \citep{DeHaas2005}. Besides economic migration, another approach prevalent in the literature is lifestyle migration \citep{BensonOReilly2009,BensonOReilly2016,SaarSaar2020}. Evidence suggests that the two motivations cannot be wholly separated \citep{Bobek2020,Croucher2015, MaileGriffiths2012}.

To summarise, people often choose with their foot between countries. Someone migrates from country $A$ to country $B$ if the latter is judged to be a better place. Data on international migration is highly unreliable, most countries do not collect data on their leaving citizens. Because of their importance to labour-exporting countries, the most visible aspect of the international migration is the amount of remittances received \citep{Adams2003}. A remittance is a transfer of money by a foreign worker to an individual in their home country. It constitutes a significant part of international capital flows, especially for labour-exporting countries. That is why we use these data to proxy the choice of people between countries. Using personal remittances to rank countries is not unique (see, for example, \href{https://www.indexmundi.com/facts/indicators/BX.TRF.PWKR.CD.DT/rankings}{IndexMundi}). 

Our \href{http://www.worldbank.org/en/topic/migrationremittancesdiasporaissues/brief/migration-remittances-data}{dataset} contains estimates of bilateral remittances by the World Bank, based on migrant stocks, host country incomes, and origin country incomes \citep{WorldBank2017}. They are not officially reported data since bilateral remittance flows are not registered appropriately. The estimation uses the methodology of \citet{RathaShaw2007} who have devised a simple formula to allocate the recorded remittances received by each country to the source countries. This applies a remittance function assuming that the amount sent by an average worker increases with the migrant's income but at a decreasing rate. Furthermore, in the case of migration to a country where the per capita income is lower than in the host country, the transfer is supposed to be at least as much as the per capita income of the origin country. 

Our paper contributes to the literature on alternative quality of life country indexes, surveyed in \citet{SomarribaPena2009}, but offers a new approach using pairwise comparisons.

The paper is structured as follows. Section~\ref{literature} surveys previous literature. Section~\ref{Sec2} outlines the theoretical background of our calculations. The alternative quality of life rankings are presented in Section~\ref{Sec3}, while Section~\ref{Sec4} summarises the main findings.

\section{Literature review} \label{literature}
In the recent decades, scientists have suggested several alternative approaches to define and measure the quality of life.  \citet{Brock1993} distinguish three different theories: the hedonist, preference satisfaction, and ideal theories of a good life. In the interpretation of \citet{DienerSuh1997}, the first one is an experience of the individual, which is mainly associated with subjective well-being. Preference satisfaction means that people choose those things that most enhance their quality of life within the constraints of the resources they possess. The third approach describes ``characteristics of the good life that are dictated by normative ideals based on a religious, philosophical, or other systems'' \citep[p.~189]{DienerSuh1997}.

\subsection{Measuring quality of life}
Quality of life can be defined in many ways, which makes its measurement difficult. For \citet{Liu1976}, the quality of life consists of five components: economic, political, environmental, social, health and educational.  According to \citet{BoyerSavageau1981}, however, it has nine components: climate, housing, health, crime, transportation, recreation, art, economics, and education. The index of \citet{Johnston1988} focuses on the changes instead of indicator values. The author uses twenty-one variables in nine major areas: health, public safety, education, employment, earnings and income, poverty, housing, family stability, equality. 

In his pioneering work, \citet{Morris1979} uses the concept of well-being to rank countries by the Physical Quality of Life Index (PQLI). This has been constructed the index by using life expectancy at age one, infant mortality, and adult literacy rates, taking a simple average of the three
constituents. \citet{Ram1982} improves PQLI by applying principal component analysis to determine the weights and adding per capita GNP to the variables. \citet{Diener1995} proposes the Basic QOL Index, designed primarily to discriminate between developing countries by including seven variables:
purchasing power, homicide rate, fulfillment of basic physical needs, suicide rate, literacy rate, gross human rights violations, and deforestation. The Advanced QOL Index, designed primarily for highly industrialized nations, involves also seven variables: physicians per capita, savings rate, per capita income, subjective well-being, percent attending college, income equality, and environmental treaties. Combining the two indices leds to
the Combined QOL Index \citep{Diener1995}.

The United Nation's Human Development Index (HDI) is perhaps the best known and probably the most researched measure of human development. It is a composite index of three dimensions: health (measured by life expectancy at birth), education (mean of years of schooling for adults aged 25 years and more, as well as the expected years of schooling for children of school entering age), and standard of living (gross national income per capita, previously GDP). The scores for the three variables are aggregated into a composite index using geometric mean.\footnote{~\href{http://hdr.undp.org/en/content/human-development-index-hdi}{http://hdr.undp.org/en/content/human-development-index-hdi}} The motivation behind the structure of the index is expressed in the 1990 Human Development Report as follows: ``Human development is a process of enlarging people's choices. In principle, these choices can be infinite and change over time. But at all levels of development, the three essential ones are for people to lead a long and healthy life, to acquire knowledge and to have access to resources needed for a decent standard of living \citep{UNDP1990}.'' The methodology has changed several times, see \citep{KlugmanRodriguezChoi2011} for a summary. Despite its comprehensive use, HDI has also got serious criticism \citep{Ravallion2012b}. \citet{Neumayer2001} provides a comprehensive overview  on alternative computational methods for calculating the HDI.

One of the main question of composite indices is their weighting system. Multivariate techniques present an empirical and relatively objective option for weight selection. According to \citet{Booysen2002}, principal components analysis (see \citet{GreylingTregenna2017})  and factor analysis are the most frequently used techniques. Data envelopment analysis (DEA) is also quite often applied for this purpose \citep{Despotis2005, Cherchyeetal2008, VanpuyenbroeckRogge2020}. \citet{SomarribaPena2009} distinguish three decision-making methodologies to obtain synthetic indicators in the area of welfare and quality of life for ranking European countries: principal components analysis, data envelopment analysis (DEA) and a measure of distance.  \citet{BerengerVerdier2007} propose totally fuzzy analysis and the factorial analysis of correspondences to determine weight for  standard
of living and  quality of life measures. \citet{OmraniAlizadehAmini2020} apply two various multi-criteria decision making  techniques, BWM and MULTIMOORA, to calculate a semi-HDI index for the provinces of Iran. \citet{KaragiannisKaragiannis2020} find the solution on the basis of Shannon entropy.

To conclude, there is clearly no perfect measure for such a complex concept as the quality of life. Hopefully, our approach might also be able to grab an important component of this notion.
\subsection{The significance of country rankings}
Country rankings o economic and political guides. For example, 115 countries have implemented 294 business regulatory reforms between 2018 and 2019 across the ten areas measured by Doing Business.\footnote{~\href{https://www.doingbusiness.org/en/reforms}{https://www.doingbusiness.org/en/reforms}} 

At the same time,  uncertain ranking promote rank-seeking behaviour by certain institutions. In 2007, the Malaysian Industrial Development Authority insisted that Malaysia aims to move from the 24th position to the top 10, and Kyrgyzstan expressed its wish to be among top 20 countries in the World Bank's Doing Business ranking. However, this behaviour does not reflect real economic performance, just improvement in the right indicator \citep{HoylandMoeneWillumsen2012}. 

On the other hand, the Ease of Doing Business (EDBI) has inspired structural economic reforms in Russia \citep{BroomeHomolarKranke2018}. This index is also widely used by multinationals in their investment location strategies \citep{PinheiroAlvesZambujalOliveira2012}. Finally, the European Development Fund has explicitly taken components of countries' HDI scores into account for the purposes of allocating aid \citep{DavisKingsburyMerry2012}.

\subsection{The economic effects of remittances}

Inflows of workers' remittances have been proliferating in many developing countries. Consequently, several papers conclude that remittances help reducing poverty. In their survey of 538 estimates published in 95 studies, \citet{CazacheviciHavranekHorvath2020} find that 40\% of the studies report a positive effect, 40\% report no effect, and 20\% report a negative effect.

On the one hand, remittances reduce inequality and can serve as a substitute for financial development \citep{AdamsPage2005,Inoue2018}. According to \citet{GiulianoRuizArranz2009} remittances boost growth in countries with less developed financial systems by providing an alternative way to finance investment and helping overcome liquidity constraints. 

Another approach argues that remittance is a compensatory transfer and should  negatively correlate  with GDP growth \citep{ChamiFullenkampJahjah2005}. In this case, it serves as an insurance \citep{YangChoi2007}.

Nonetheless remittances can have adverse effects, especially by leading to the Dutch disease, development in one specific sector only, while a decline in  others. \citet{AcostaLarteyMandelman2009} find that an increase in remittance flows results in a decline in labor supply and an increase in consumption demand that is biased toward non-tradables in El Salvador. \citet{RoyDixon2016} cannot reject that remittances promote an appreciation of the real exchange rate and thereby hurt the competitiveness of the tradeable sector in South Asia. \citet{AbdihChamiDagherMontiel2012} show how an increase in remittance inflows correlate with lower indices of control of corruption and government effectiveness. \citet{BerdievKimChang2013} find that remittances increase corruption, especially in non-OECD countries. However, \citet{Tyburski2012} argues that remittances mitigate corruption by increasing government accountability and providing other incentives to reform. 

\section{Methodology} \label{Sec2}

We consider one unit of money transferred from a country to another country such that the former is preferred over the latter by one voter, and use techniques from social choice theory to evaluate these ``votes''.
Let us introduce the matrix $\mathbf{A} = \left[ a_{ij} \right] \in \mathbb{R}^{n \times n}$ of bilateral remittances among $n$ countries, where $a_{ij}$ is the sum of transfers from country $i$ to country $j$ in the given period.
It immediately determines the skew-symmetric \emph{results matrix} $\mathbf{R} = \mathbf{A} - \mathbf{A}^\top$ and the symmetric \emph{matches matrix} $\mathbf{M} = \mathbf{A} + \mathbf{A}^\top$.

Perhaps the simplest measure is to calculate the net remittances (the difference of total outflow and total inflow), denote it by $s_i = \sum_{j=1}^n r_{ij}$ for each country $i$.
Dividing these amounts by the total remittance flow (the sum of total outflow and total inflow) $\sum_{j=1}^n m_{ij}$ of the country leads to $p_i$.

Finally, the least squares method adjusts net remittances by taking the whole network of  flows into account.
The least squares weight $q_i$ of country $i$ can be obtained as the solution of the following optimization problem:

\begin{equation}\label{eq1}
\min_{q \in \mathbb{R}^n} \sum_{1 \leq i,j \leq n} m_{ij} \left( \frac{r_{ij}}{m_{ij}} - q_i + q_j \right)^2.
\end{equation}

The first order conditions of optimality lead to a linear equation for each country $i$:
\begin{equation}\label{eq2}
\left( \sum_{j=1}^n  m_{ij} \right) q_i  - \sum_{j=1}^n m_{ij} q_j = s_i = \sum_{j=1}^n r_{ij}.
\end{equation}
After normalizing the weights by $\sum_{i=1}^n q_i = 0$, the solution of this system becomes unique if the countries are connected at least indirectly by transfers, that is, the multigraph of bilateral remittances is connected \citep{CaklovicKurdija2017}. Our dataset has satisfied this condition in every year.

All of the weight vectors above determine a ranking $\succeq$ of the countries. For example, the ranking from the least squares method is given by $q_i \geq q_j \iff i \succeq j$.

In order to highlight the characteristics of the three methods, two axiomatic properties have been considered.
First, the ranking is required to be \emph{invariant to country sizes}, that is, countries $i$ and $j$ should get the same rank if country $j$ has a fixed proportion of transfers to and from every third country as country $i$. This axiom has been introduced in \citet{CsatoToth2020} under the name \emph{size invariance}.

Second, it should be \emph{independent of bridge country}, namely, in a hypothetical world consisting of two set of countries connected only by a particular country called bridge country, the relative rankings within each set of countries are should be independent of the remmittances among the countries in the other set \citep{Gonzalez-DiazHendrickxLohmann2014}.

Ranking by net remittances violates both properties. The ranking derived from vector $\mathbf{p}$ is invariant to country sizes but does not meet bridge country independence. The least squares method satisfies both axioms, therefore we suggest to apply this procedure.

Note that the least squares method is equivalent to the Potential Method \citep{CaklovicKurdija2017}, to the EKS (\'Eltet\H{o}--K\"oves--Szulc) method used for international price comparisons by the OECD \citep{EltetoKoves1964, Szulc1964}, and to the Logarithmic Least Squares Method defined in the framework of (incomplete) multiplicative pairwise comparison matrices \citep{BozokiFulopRonyai2010, BozokiTsyganok2019}.

%Its axiomatic properties are discussed by \citet{Gonzalez-DiazHendrickxLohmann2014}, while \citet{Csato2015a} gives a graph interpretation of the least squares method.
%\citet{Csato2018c} and \citet{Csato2019a} provide characterizations of the procedure on a restricted domain. Finally, \citet{CsatoRonyai2016} present a potential failure of it, but \citet{Csato2018f} and \citet{Csato2018g} prove that it is impossible to find a perfect ranking method in the case of such complex problems.

%The method has been recently applied for ranking the participating countries of the Eurovision Song Contest \citep{CaklovicKurdija2017}, as well as for the comparison of top historical players in Go \citep{ChaoKouLiPeng2018} and tennis \citep{BozokiCsatoTemesi2016}.

It has been recently applied for ranking the teams in a Swiss system chess team tournament \citep{Csato2013a, Csato2017c}, the Hungarian universities on the basis of applicants' preferences \citep{CsatoToth2020}, the participating countries of the Eurovision Song Contest \citep{CaklovicKurdija2017}, as well as for the comparison of top historical players in Go \citep{ChaoKouLiPeng2018} and tennis \citep{BozokiCsatoTemesi2016}.

As an illustration, consider the following example with four countries.
\begin{figure}[!htbp]
\caption{Remittances between the four countries in Example~\ref{Examp1}}
\label{Fig1}
\centering
\begin{tikzpicture}[scale=0.8, auto=center, transform shape, >=triangle 45, ->, >=stealth', semithick, shorten >=1pt]
  \tikzstyle{every state}=[fill=white,draw=black,text=black]
  
  \node[state] (A) at (150:4) {$A$};
  \node[state] (B) at (210:4) {$B$};
  \node[state] (C) at (0,0)   {$C$};
  \node[state] (D) at (0:4)   {$D$}; 

  \path (A) edge    [bend left=15, line width=0.45mm]        node [above, near start] {15} (C)
        (B) edge 	[bend left=15, line width=0.9mm]		 node [above, near start] {30} (C)
        (C) edge    [bend left=15, line width=0.3mm]         node [below] {10} (B)
            edge    [bend left=15, line width=0.15mm]        node [below, near end] {5}  (A)
            edge    [bend left=15, line width=0.3mm]         node [above] {10} (D)
        (D) edge    [bend left=15, line width=0.3mm]         node [below] {10} (C);
\end{tikzpicture}
\end{figure}

\begin{example} \label{Examp1}
Consider the four countries shown in Figure~\ref{Fig1}, where the directed edges represent the remittances, and their weights correspond to the amount of the transfer.
The remittances, results, and matches matrices are as follows:
\[
\mathbf{A} = \left[
\begin{array}{K{2em} K{2em} K{2em} K{2em} }
 0 & 0 & 15 & 0 \\  
 0 & 0 & 30 & 0 \\  
 5 & 10 & 0 & 10 \\  
 0 & 0 & 10 & 0 \\
\end{array}
\right]
\text{, }
\mathbf{R} = \left[
\begin{array}{K{2em} K{2em} K{2em} K{2em}}
  0 & 0 & 10 & 0 \\ 
  0 & 0 & 20 & 0 \\  
  -10 & -20 & 0 & 0 \\  
  0 & 0 & 0 & 0 \\
\end{array}
\right]
\text{, and }
\]

\[
\mathbf{M} = \left[
\begin{array}{K{2em} K{2em} K{2em} K{2em}}
  0 & 0 & 20 & 0 \\ 
  0 & 0 & 40 & 0 \\ 
 20 & 40 & 0 & 20 \\  
 0 & 0 & 20 & 0 \\
\end{array}
\right].
\]

Remittances to and from country $B$ are proportional to the remittances to and from country $A$, thus size invariance implies the same rank for these countries.
$C$ is a bridge country between the sets $\{ A; B; C \}$ and $\{ C; D \}$, that is, the relative ranking within the second set is determined only by the transfers between countries $C$ and $D$, and it makes no sense to rank one of them above the other.
\begin{table}[!tbp]
\caption{The weight vectors of Example~\ref{Examp1}}
\label{Table1}
\centering
\begin{tabularx}{0.4\textwidth}{cRRR} \toprule
Country & $\mathbf{s}(\mathbf{A})$ & $\mathbf{p}(\mathbf{A})$ & $\mathbf{q}(\mathbf{A})$  \\ \midrule
$A$ & $10$  & $1/2$    & $1/4$  \\ 
$B$ & $20$  & $1/2$    & $1/4$  \\ 
$C$ & $-30$ & $-3/8$ & $-1/4$ \\  
$D$ & $0$   & $0$      & $-1/4$ \\ \bottomrule
\end{tabularx} 
\end{table}

The weights according to the three methods are shown in Table~\ref{Table1}.
The vector $\mathbf{s}$ of net remittances violates the two properties suggested above, the ratio of the net and total remittances ($\mathbf{p}$) satisfies size invariance, but does not meet bridge country independence, while the least squares method ($\mathbf{q}$) meets both requirements.
\end{example}

\section{Alternative quality of life rankings} \label{Sec3}

In the following, the results of our calculations with the methodology suggested in Section~\ref{Sec2} are presented.

\subsection{The ranking in 2015} \label{Subsec31}

\begin{table}[!htbp]
\caption{Ranking of European countries on the basis of remittances in 2015}
\label{Table2}
\centering
    \begin{tabularx}{\linewidth}{cCCCCCCC} \toprule
Country & $\mathbf{s}(\mathbf{A})$ & $\mathbf{p}(\mathbf{A})$ & $\mathbf{q}(\mathbf{A})$ & Country & $\mathbf{s}(\mathbf{A})$ & $\mathbf{p}(\mathbf{A})$ & $\mathbf{q}(\mathbf{A})$ \\ \hline 

    AL    & \cellcolor[rgb]{ .635,  .796,  .612}25 & \cellcolor[rgb]{ .906,  .949,  .902}37 & \cellcolor[rgb]{ .863,  .922,  .855}35 & IS    & \cellcolor[rgb]{ .463,  .698,  .427}18 & \cellcolor[rgb]{ .533,  .737,  .502}21 & \cellcolor[rgb]{ .486,  .71,  .455}19 \\
    AT    & \cellcolor[rgb]{ .294,  .604,  .251}10 & \cellcolor[rgb]{ .361,  .643,  .322}13 & \cellcolor[rgb]{ .431,  .678,  .396}16 & IT    & \cellcolor[rgb]{ .157,  .529,  .106}5 & \cellcolor[rgb]{ .251,  .58,  .208}9 & \cellcolor[rgb]{ .251,  .58,  .208}9 \\
    BA    & \cellcolor[rgb]{ .792,  .882,  .78}32 & 41    & 41    & LT    & \cellcolor[rgb]{ .694,  .827,  .678}28 & \cellcolor[rgb]{ .953,  .973,  .949}39 & \cellcolor[rgb]{ .765,  .867,  .749}31 \\
    BE    & \cellcolor[rgb]{ .953,  .973,  .949}39 & \cellcolor[rgb]{ .612,  .78,  .588}24 & \cellcolor[rgb]{ .565,  .757,  .541}22 & LU    & \cellcolor[rgb]{ .576,  .765,  .553}23 & \cellcolor[rgb]{ .486,  .71,  .455}19 & \cellcolor[rgb]{ .533,  .737,  .502}21 \\
    BG    & \cellcolor[rgb]{ .749,  .859,  .733}30 & \cellcolor[rgb]{ .976,  .984,  .973}40 & \cellcolor[rgb]{ .953,  .973,  .949}39 & LV    & \cellcolor[rgb]{ .671,  .816,  .651}27 & \cellcolor[rgb]{ .835,  .906,  .824}34 & \cellcolor[rgb]{ .694,  .827,  .678}28 \\
    BY    & \cellcolor[rgb]{ .384,  .655,  .345}14 & \cellcolor[rgb]{ .384,  .655,  .345}14 & \cellcolor[rgb]{ .361,  .643,  .322}13 & MD    & \cellcolor[rgb]{ .718,  .843,  .702}29 & \cellcolor[rgb]{ .859,  .922,  .851}35 & \cellcolor[rgb]{ .812,  .894,  .8}33 \\
    CH    & \cellcolor[rgb]{ .157,  .525,  .106}4 & \cellcolor[rgb]{ .133,  .514,  .082}3 & \cellcolor[rgb]{ .11,  .502,  .059}2 & ME    & \cellcolor[rgb]{ .553,  .749,  .529}22 & \cellcolor[rgb]{ .929,  .961,  .925}38 & \cellcolor[rgb]{ .976,  .984,  .973}40 \\
    CY    & \cellcolor[rgb]{ .361,  .643,  .322}13 & \cellcolor[rgb]{ .247,  .576,  .204}8 & \cellcolor[rgb]{ .204,  .553,  .153}6 & MK    & \cellcolor[rgb]{ .51,  .725,  .478}20 & \cellcolor[rgb]{ .694,  .827,  .678}28 & \cellcolor[rgb]{ .741,  .855,  .725}30 \\
    CZ    & \cellcolor[rgb]{ .612,  .78,  .588}24 & \cellcolor[rgb]{ .522,  .729,  .49}20 & \cellcolor[rgb]{ .635,  .796,  .612}25 & MT    & \cellcolor[rgb]{ .486,  .71,  .455}19 & \cellcolor[rgb]{ .671,  .816,  .651}27 & \cellcolor[rgb]{ .576,  .765,  .553}23 \\
    DE    & \cellcolor[rgb]{ .133,  .514,  .082}3 & \cellcolor[rgb]{ .318,  .616,  .275}11 & \cellcolor[rgb]{ .337,  .627,  .298}12 & NL    & \cellcolor[rgb]{ .204,  .553,  .157}7 & \cellcolor[rgb]{ .09,  .49,  .035}2 & \cellcolor[rgb]{ .157,  .529,  .106}5 \\
    DK    & \cellcolor[rgb]{ .337,  .627,  .298}12 & \cellcolor[rgb]{ .337,  .627,  .298}12 & \cellcolor[rgb]{ .294,  .604,  .251}10 & NO    & \cellcolor[rgb]{ .227,  .569,  .18}8 & \cellcolor[rgb]{ .133,  .514,  .082}4 & \cellcolor[rgb]{ .133,  .514,  .082}4 \\
    EE    & \cellcolor[rgb]{ .545,  .745,  .518}21 & \cellcolor[rgb]{ .725,  .847,  .71}29 & \cellcolor[rgb]{ .678,  .82,  .659}27 & PL    & \cellcolor[rgb]{ .976,  .984,  .973}40 & \cellcolor[rgb]{ .812,  .894,  .8}33 & \cellcolor[rgb]{ .718,  .843,  .702}29 \\
    ES    & \cellcolor[rgb]{ .204,  .553,  .153}6 & \cellcolor[rgb]{ .294,  .604,  .251}10 & \cellcolor[rgb]{ .318,  .616,  .275}11 & PT    & \cellcolor[rgb]{ .835,  .906,  .824}34 & \cellcolor[rgb]{ .647,  .804,  .627}26 & \cellcolor[rgb]{ .6,  .776,  .576}24 \\
    FI    & \cellcolor[rgb]{ .431,  .678,  .396}16 & \cellcolor[rgb]{ .408,  .667,  .373}15 & \cellcolor[rgb]{ .384,  .655,  .345}14 & RO    & \cellcolor[rgb]{ .882,  .933,  .875}36 & \cellcolor[rgb]{ .882,  .933,  .875}36 & \cellcolor[rgb]{ .835,  .906,  .824}34 \\
    FR    & \cellcolor[rgb]{ .906,  .949,  .902}37 & \cellcolor[rgb]{ .475,  .706,  .443}18 & \cellcolor[rgb]{ .451,  .694,  .42}17 & RS    & \cellcolor[rgb]{ .859,  .922,  .851}35 & \cellcolor[rgb]{ .741,  .855,  .725}30 & \cellcolor[rgb]{ .906,  .945,  .898}37 \\
    GB    & \cellcolor[rgb]{ .09,  .49,  .035}1 & \cellcolor[rgb]{ .09,  .49,  .035}1 & \cellcolor[rgb]{ .09,  .49,  .035}1 & RU    & \cellcolor[rgb]{ .09,  .49,  .035}2 & \cellcolor[rgb]{ .204,  .553,  .157}7 & \cellcolor[rgb]{ .227,  .569,  .18}8 \\
    GR    & \cellcolor[rgb]{ .318,  .616,  .275}11 & \cellcolor[rgb]{ .204,  .553,  .153}6 & \cellcolor[rgb]{ .224,  .565,  .176}7 & SE    & \cellcolor[rgb]{ .392,  .659,  .357}15 & \cellcolor[rgb]{ .416,  .671,  .38}16 & \cellcolor[rgb]{ .392,  .659,  .357}15 \\
    HR    & \cellcolor[rgb]{ .659,  .808,  .635}26 & \cellcolor[rgb]{ .635,  .796,  .612}25 & \cellcolor[rgb]{ .929,  .961,  .925}38 & SI    & \cellcolor[rgb]{ .439,  .686,  .404}17 & \cellcolor[rgb]{ .439,  .686,  .404}17 & \cellcolor[rgb]{ .647,  .804,  .627}26 \\
    HU    & \cellcolor[rgb]{ .929,  .961,  .925}38 & \cellcolor[rgb]{ .792,  .882,  .78}32 & \cellcolor[rgb]{ .792,  .882,  .78}32 & SK    & \cellcolor[rgb]{ .765,  .867,  .749}31 & \cellcolor[rgb]{ .765,  .867,  .749}31 & \cellcolor[rgb]{ .882,  .933,  .875}36 \\
    IE    & \cellcolor[rgb]{ .271,  .592,  .227}9 & \cellcolor[rgb]{ .18,  .541,  .129}5 & \cellcolor[rgb]{ .133,  .514,  .082}3 & UA    & \cellcolor[rgb]{ .812,  .894,  .8}33 & \cellcolor[rgb]{ .553,  .749,  .529}22 & \cellcolor[rgb]{ .51,  .725,  .478}20 \\ \hline
          &       &       &       & Other & 41    & \cellcolor[rgb]{ .576,  .765,  .553}23 & \cellcolor[rgb]{ .463,  .698,  .427}18 \\

\hline
    \end{tabularx}
\end{table}

Due to the lack of reliable data for some developing countries, we have restricted the investigation to Europe.  To select European countries, the United Nations geoscheme has been used. However, we have omitted Andorra, Liechtenstein, Monaco, San Marino, and Vatican City because lack of data, but we have enrolled Cyprus due to its membership of the European Union. The value of remittances can be found in Table~\ref{netrem} in the Appendix.

Table~\ref{Table2} presents the ranking of the countries with the three methods based on the transfers in 2015 such that all non-European countries are regarded as one entity. ISO 3166-1 standard alpha-2 codes are used to abbreviate countries, which can be found in Table~\ref{table5} in the Appendix. Lighter colour indicates a worse rank. The largest difference between the ranking from $\mathbf{s}$ and $\mathbf{q}$ is in the case of France. Net remittances place it to the 37th position, while the least squares method gives the 17th rank. The reason is the size effect: France is one of the largest countries in Europe, hence it is natural that both inflow and outflow  are huge (more than 20 billion USD), which implicates that the difference is also great ($-2482$ million USD). In this case, even Albania overtakes France with net remittances of $-852$ million USD, but with almost five times higher inflow than outflow.

It can be realized from formula~\eqref{eq2} that $q_i$ is close to the size-invariant ratio $p_i=s_i/(\sum_{j=1}^n m_{ij})$  if the weights of the countries connected to country $i$ by remittances are close to the average weight of 0. On the other hand, $q_i$ becomes higher (lower) than this ratio if country $i$ is mainly connected to higher (lower) ranked countries by the transfers.
This adjustment is the largest in the case of Croatia, Lithuania, and Slovenia. The main destinations for Croatia are Germany and Serbia, while it receives workers from Bosnia and Herzegovina, Serbia, and Slovenia, therefore Croatia is mainly connected to lower ranked countries.
On the other hand, Lithuania (LT) is predominantly connected to some higher ranked countries (the United Kingdom, Russia), which implicates its better rank with $q_i$.

\begin{table}[!htbp]
\caption{Quality of life rankings by the least squares method}
\label{Table3}
\centering
\centerline{
\begin{tabular}{cccccccccccccc} \toprule
Country & 2010  & 2011  & 2012  & 2013  & 2014  & 2015  & Country & 2010  & 2011  & 2012  & 2013  & 2014  & 2015 \\ \hline
    AL    & \cellcolor[rgb]{ .929,  .961,  .925}38 & \cellcolor[rgb]{ .929,  .961,  .925}38 & \cellcolor[rgb]{ .906,  .949,  .902}37 & \cellcolor[rgb]{ .906,  .949,  .902}37 & \cellcolor[rgb]{ .886,  .933,  .878}36 & \cellcolor[rgb]{ .863,  .922,  .855}35 & IS    & \cellcolor[rgb]{ .09,  .49,  .035}1 & \cellcolor[rgb]{ .09,  .49,  .035}1 & \cellcolor[rgb]{ .11,  .502,  .059}2 & \cellcolor[rgb]{ .51,  .725,  .478}19 & \cellcolor[rgb]{ .553,  .749,  .529}21 & \cellcolor[rgb]{ .51,  .725,  .478}19 \\
    AT    & \cellcolor[rgb]{ .475,  .706,  .443}18 & \cellcolor[rgb]{ .475,  .706,  .443}18 & \cellcolor[rgb]{ .522,  .729,  .49}20 & \cellcolor[rgb]{ .337,  .627,  .298}12 & \cellcolor[rgb]{ .361,  .643,  .322}13 & \cellcolor[rgb]{ .431,  .678,  .396}16 & IT    & \cellcolor[rgb]{ .322,  .62,  .282}11 & \cellcolor[rgb]{ .298,  .608,  .255}10 & \cellcolor[rgb]{ .345,  .631,  .306}12 & \cellcolor[rgb]{ .251,  .58,  .208}8 & \cellcolor[rgb]{ .251,  .58,  .208}8 & \cellcolor[rgb]{ .275,  .592,  .231}9 \\
    BA    & 41    & 41    & 41    & 41    & 41    & 41    & LT    & \cellcolor[rgb]{ .906,  .945,  .898}36 & \cellcolor[rgb]{ .906,  .945,  .898}36 & \cellcolor[rgb]{ .882,  .933,  .875}35 & \cellcolor[rgb]{ .812,  .894,  .8}32 & \cellcolor[rgb]{ .788,  .882,  .776}31 & \cellcolor[rgb]{ .788,  .882,  .776}31 \\
    BE    & \cellcolor[rgb]{ .565,  .757,  .541}22 & \cellcolor[rgb]{ .565,  .757,  .541}22 & \cellcolor[rgb]{ .565,  .757,  .541}22 & \cellcolor[rgb]{ .659,  .808,  .635}26 & \cellcolor[rgb]{ .635,  .796,  .612}25 & \cellcolor[rgb]{ .565,  .757,  .541}22 & LU    & \cellcolor[rgb]{ .6,  .776,  .576}23 & \cellcolor[rgb]{ .6,  .776,  .576}23 & \cellcolor[rgb]{ .6,  .776,  .576}23 & \cellcolor[rgb]{ .6,  .776,  .576}23 & \cellcolor[rgb]{ .6,  .776,  .576}23 & \cellcolor[rgb]{ .553,  .749,  .529}21 \\
    BG    & \cellcolor[rgb]{ .953,  .973,  .949}39 & \cellcolor[rgb]{ .953,  .973,  .949}39 & \cellcolor[rgb]{ .953,  .973,  .949}39 & \cellcolor[rgb]{ .953,  .973,  .949}39 & \cellcolor[rgb]{ .953,  .973,  .949}39 & \cellcolor[rgb]{ .953,  .973,  .949}39 & LV    & \cellcolor[rgb]{ .624,  .788,  .604}24 & \cellcolor[rgb]{ .624,  .788,  .604}24 & \cellcolor[rgb]{ .647,  .804,  .627}25 & \cellcolor[rgb]{ .533,  .737,  .502}20 & \cellcolor[rgb]{ .576,  .765,  .553}22 & \cellcolor[rgb]{ .718,  .843,  .702}28 \\
    BY    & \cellcolor[rgb]{ .408,  .667,  .373}15 & \cellcolor[rgb]{ .408,  .667,  .373}15 & \cellcolor[rgb]{ .408,  .667,  .373}15 & \cellcolor[rgb]{ .408,  .667,  .373}15 & \cellcolor[rgb]{ .384,  .655,  .345}14 & \cellcolor[rgb]{ .361,  .643,  .322}13 & MD    & \cellcolor[rgb]{ .859,  .922,  .851}34 & \cellcolor[rgb]{ .859,  .922,  .851}34 & \cellcolor[rgb]{ .859,  .922,  .851}34 & \cellcolor[rgb]{ .788,  .882,  .776}31 & \cellcolor[rgb]{ .765,  .867,  .749}30 & \cellcolor[rgb]{ .835,  .906,  .824}33 \\
    CH    & \cellcolor[rgb]{ .247,  .576,  .204}8 & \cellcolor[rgb]{ .247,  .576,  .204}8 & \cellcolor[rgb]{ .247,  .576,  .204}8 & \cellcolor[rgb]{ .133,  .514,  .082}3 & \cellcolor[rgb]{ .11,  .502,  .059}2 & \cellcolor[rgb]{ .11,  .502,  .059}2 & ME    & \cellcolor[rgb]{ .929,  .961,  .925}37 & \cellcolor[rgb]{ .929,  .961,  .925}37 & \cellcolor[rgb]{ .953,  .973,  .949}38 & 40    & 40    & 40 \\
    CY    & \cellcolor[rgb]{ .18,  .541,  .129}5 & \cellcolor[rgb]{ .157,  .525,  .106}4 & \cellcolor[rgb]{ .157,  .525,  .106}4 & \cellcolor[rgb]{ .11,  .502,  .059}2 & \cellcolor[rgb]{ .224,  .565,  .176}7 & \cellcolor[rgb]{ .204,  .553,  .153}6 & MK    & \cellcolor[rgb]{ .812,  .894,  .8}32 & \cellcolor[rgb]{ .812,  .894,  .8}32 & \cellcolor[rgb]{ .788,  .882,  .776}31 & \cellcolor[rgb]{ .906,  .945,  .898}36 & \cellcolor[rgb]{ .882,  .933,  .875}35 & \cellcolor[rgb]{ .765,  .867,  .749}30 \\
    CZ    & \cellcolor[rgb]{ .678,  .82,  .659}27 & \cellcolor[rgb]{ .659,  .808,  .635}26 & \cellcolor[rgb]{ .659,  .808,  .635}26 & \cellcolor[rgb]{ .545,  .745,  .518}21 & \cellcolor[rgb]{ .522,  .729,  .49}20 & \cellcolor[rgb]{ .635,  .796,  .612}25 & MT    & \cellcolor[rgb]{ .51,  .725,  .478}19 & \cellcolor[rgb]{ .51,  .725,  .478}19 & \cellcolor[rgb]{ .486,  .71,  .455}18 & \cellcolor[rgb]{ .741,  .855,  .725}29 & \cellcolor[rgb]{ .859,  .922,  .851}34 & \cellcolor[rgb]{ .6,  .776,  .576}23 \\
    DE    & \cellcolor[rgb]{ .294,  .604,  .251}10 & \cellcolor[rgb]{ .318,  .616,  .275}11 & \cellcolor[rgb]{ .318,  .616,  .275}11 & \cellcolor[rgb]{ .318,  .616,  .275}11 & \cellcolor[rgb]{ .318,  .616,  .275}11 & \cellcolor[rgb]{ .337,  .627,  .298}12 & NL    & \cellcolor[rgb]{ .227,  .569,  .18}7 & \cellcolor[rgb]{ .227,  .569,  .18}7 & \cellcolor[rgb]{ .204,  .553,  .157}6 & \cellcolor[rgb]{ .18,  .541,  .133}5 & \cellcolor[rgb]{ .133,  .514,  .082}3 & \cellcolor[rgb]{ .18,  .541,  .133}5 \\
    DK    & \cellcolor[rgb]{ .337,  .627,  .298}12 & \cellcolor[rgb]{ .337,  .627,  .298}12 & \cellcolor[rgb]{ .361,  .643,  .322}13 & \cellcolor[rgb]{ .294,  .604,  .251}10 & \cellcolor[rgb]{ .294,  .604,  .251}10 & \cellcolor[rgb]{ .294,  .604,  .251}10 & NO    & \cellcolor[rgb]{ .204,  .553,  .157}6 & \cellcolor[rgb]{ .204,  .553,  .157}6 & \cellcolor[rgb]{ .227,  .569,  .18}7 & \cellcolor[rgb]{ .227,  .569,  .18}7 & \cellcolor[rgb]{ .18,  .541,  .133}5 & \cellcolor[rgb]{ .157,  .529,  .106}4 \\
    EE    & \cellcolor[rgb]{ .659,  .808,  .635}26 & \cellcolor[rgb]{ .678,  .82,  .659}27 & \cellcolor[rgb]{ .678,  .82,  .659}27 & \cellcolor[rgb]{ .635,  .796,  .612}25 & \cellcolor[rgb]{ .678,  .82,  .659}27 & \cellcolor[rgb]{ .678,  .82,  .659}27 & PL    & \cellcolor[rgb]{ .718,  .843,  .702}28 & \cellcolor[rgb]{ .718,  .843,  .702}28 & \cellcolor[rgb]{ .718,  .843,  .702}28 & \cellcolor[rgb]{ .718,  .843,  .702}28 & \cellcolor[rgb]{ .718,  .843,  .702}28 & \cellcolor[rgb]{ .741,  .855,  .725}29 \\
    ES    & \cellcolor[rgb]{ .271,  .592,  .227}9 & \cellcolor[rgb]{ .271,  .592,  .227}9 & \cellcolor[rgb]{ .271,  .592,  .227}9 & \cellcolor[rgb]{ .271,  .592,  .227}9 & \cellcolor[rgb]{ .271,  .592,  .227}9 & \cellcolor[rgb]{ .318,  .616,  .275}11 & PT    & \cellcolor[rgb]{ .533,  .737,  .502}20 & \cellcolor[rgb]{ .553,  .749,  .529}21 & \cellcolor[rgb]{ .553,  .749,  .529}21 & \cellcolor[rgb]{ .694,  .827,  .678}27 & \cellcolor[rgb]{ .624,  .788,  .604}24 & \cellcolor[rgb]{ .624,  .788,  .604}24 \\
    FI    & \cellcolor[rgb]{ .451,  .694,  .42}17 & \cellcolor[rgb]{ .451,  .694,  .42}17 & \cellcolor[rgb]{ .451,  .694,  .42}17 & \cellcolor[rgb]{ .431,  .678,  .396}16 & \cellcolor[rgb]{ .475,  .706,  .443}18 & \cellcolor[rgb]{ .384,  .655,  .345}14 & RO    & \cellcolor[rgb]{ .882,  .933,  .875}35 & \cellcolor[rgb]{ .882,  .933,  .875}35 & \cellcolor[rgb]{ .906,  .945,  .898}36 & \cellcolor[rgb]{ .882,  .933,  .875}35 & \cellcolor[rgb]{ .835,  .906,  .824}33 & \cellcolor[rgb]{ .859,  .922,  .851}34 \\
    FR    & \cellcolor[rgb]{ .361,  .643,  .322}13 & \cellcolor[rgb]{ .384,  .655,  .345}14 & \cellcolor[rgb]{ .431,  .678,  .396}16 & \cellcolor[rgb]{ .451,  .694,  .42}17 & \cellcolor[rgb]{ .451,  .694,  .42}17 & \cellcolor[rgb]{ .451,  .694,  .42}17 & RS    & 40    & 40    & 40    & \cellcolor[rgb]{ .953,  .973,  .949}38 & \cellcolor[rgb]{ .953,  .973,  .949}38 & \cellcolor[rgb]{ .929,  .961,  .925}37 \\
    GB    & \cellcolor[rgb]{ .157,  .525,  .106}4 & \cellcolor[rgb]{ .18,  .541,  .129}5 & \cellcolor[rgb]{ .18,  .541,  .129}5 & \cellcolor[rgb]{ .09,  .49,  .035}1 & \cellcolor[rgb]{ .09,  .49,  .035}1 & \cellcolor[rgb]{ .09,  .49,  .035}1 & RU    & \cellcolor[rgb]{ .392,  .659,  .357}14 & \cellcolor[rgb]{ .369,  .647,  .329}13 & \cellcolor[rgb]{ .298,  .608,  .255}10 & \cellcolor[rgb]{ .204,  .553,  .157}6 & \cellcolor[rgb]{ .204,  .553,  .157}6 & \cellcolor[rgb]{ .251,  .58,  .208}8 \\
    GR    & \cellcolor[rgb]{ .545,  .745,  .518}21 & \cellcolor[rgb]{ .522,  .729,  .49}20 & \cellcolor[rgb]{ .384,  .655,  .345}14 & \cellcolor[rgb]{ .361,  .643,  .322}13 & \cellcolor[rgb]{ .337,  .627,  .298}12 & \cellcolor[rgb]{ .224,  .565,  .176}7 & SE    & \cellcolor[rgb]{ .133,  .514,  .082}3 & \cellcolor[rgb]{ .133,  .514,  .082}3 & \cellcolor[rgb]{ .133,  .514,  .082}3 & \cellcolor[rgb]{ .392,  .659,  .357}14 & \cellcolor[rgb]{ .416,  .671,  .38}15 & \cellcolor[rgb]{ .416,  .671,  .38}15 \\
    HR    & \cellcolor[rgb]{ .725,  .847,  .71}29 & \cellcolor[rgb]{ .725,  .847,  .71}29 & \cellcolor[rgb]{ .725,  .847,  .71}29 & \cellcolor[rgb]{ .749,  .859,  .733}30 & \cellcolor[rgb]{ .725,  .847,  .71}29 & \cellcolor[rgb]{ .929,  .961,  .925}38 & SI    & \cellcolor[rgb]{ .765,  .867,  .749}30 & \cellcolor[rgb]{ .788,  .882,  .776}31 & \cellcolor[rgb]{ .812,  .894,  .8}32 & \cellcolor[rgb]{ .576,  .765,  .553}22 & \cellcolor[rgb]{ .671,  .816,  .651}26 & \cellcolor[rgb]{ .671,  .816,  .651}26 \\
    HU    & \cellcolor[rgb]{ .773,  .871,  .757}31 & \cellcolor[rgb]{ .749,  .859,  .733}30 & \cellcolor[rgb]{ .749,  .859,  .733}30 & \cellcolor[rgb]{ .839,  .91,  .827}34 & \cellcolor[rgb]{ .792,  .882,  .78}32 & \cellcolor[rgb]{ .792,  .882,  .78}32 & SK    & \cellcolor[rgb]{ .835,  .906,  .824}33 & \cellcolor[rgb]{ .835,  .906,  .824}33 & \cellcolor[rgb]{ .835,  .906,  .824}33 & \cellcolor[rgb]{ .835,  .906,  .824}33 & \cellcolor[rgb]{ .929,  .961,  .925}37 & \cellcolor[rgb]{ .906,  .945,  .898}36 \\
    IE    & \cellcolor[rgb]{ .11,  .502,  .059}2 & \cellcolor[rgb]{ .11,  .502,  .059}2 & \cellcolor[rgb]{ .09,  .49,  .035}1 & \cellcolor[rgb]{ .157,  .525,  .106}4 & \cellcolor[rgb]{ .157,  .525,  .106}4 & \cellcolor[rgb]{ .133,  .514,  .082}3 & UA    & \cellcolor[rgb]{ .647,  .804,  .627}25 & \cellcolor[rgb]{ .647,  .804,  .627}25 & \cellcolor[rgb]{ .624,  .788,  .604}24 & \cellcolor[rgb]{ .624,  .788,  .604}24 & \cellcolor[rgb]{ .51,  .725,  .478}19 & \cellcolor[rgb]{ .533,  .737,  .502}20 \\ \hline
       &  &  &  &  & & & Other & \cellcolor[rgb]{ .439,  .686,  .404}16 & \cellcolor[rgb]{ .439,  .686,  .404}16 & \cellcolor[rgb]{ .51,  .725,  .478}19 & \cellcolor[rgb]{ .486,  .71,  .455}18 & \cellcolor[rgb]{ .439,  .686,  .404}16 & \cellcolor[rgb]{ .486,  .71,  .455}18 \\
 \hline
\end{tabular}
}

\end{table}

According to Table~\ref{Table3}, the least squares ranking is relatively robust across the years and does not yield many unexpected results. For example, the four members of the Visegr\'ad Group are around the 30th place, only the Czech Republic shows some improvement in the years 2013 and 2014. 

However, there are some counterintuitive findings. Data problems are responsible for the decline in the performance of Iceland and Sweden. The top position of Cyprus can be probably explained by the significant role of its banks in international finance. The United Kingdom leads the ranking in certain years partly due to its liberal migration policy. Russia gains from retaining connections of the Soviet era, as well as from the huge regional inequalities caused by the agglomerations of Moscow and Saint Petersburg. 

\subsection{Comparison with the HDI and the World Happiness Report} \label{Subsec33}

The World Happiness Report (WHR) is published by the United Nations Sustainable Development Solutions Network and ranks 156 countries according to the perceived happiness of their citizens, measured primarily with the Gallup World Poll.\footnote{~\href{https://worldhappiness.report/}{https://worldhappiness.report/}} It is a subjective well-being country ranking since 2012.
It uses a three-year range to determine the rank, thus the year 2015 refers to the order made from data  2014--16, which was published in 2017.
Table~\ref{spearman} shows the pairwise Spearman's rank order correlations between our alternative ranking, the HDI, and the WHR. There is a modestly strong positive relationship between the rankings, and all correlation is statistically significant at every reasonable level. The HDI is a somewhat closer to our ranking than the WHR. It is worth mentioning that the connection between the HDI and the WHR is quite strong for both years. 
\begin{table}[htbp]

\caption{Spearman's rank order correlation}
\label{spearman}%
  \centering
\begin{threeparttable}
\begin{tabularx}{0.8\linewidth}{l CC CCC}\toprule
&   HDI2015 & WHR2015 & LS2014 & HDI2014 & WHR2014 \\ \midrule
LS2015 &   .714* & .612* & .964* & .708* & .608* \\
HDI2015 &        & .865* & .691* & .989* & .837* \\
WHR2015 &              && .601* & .873* & .987* \\
LS2014 &             &       &  & .683* & .599* \\
HDI2014 &              &       &       &  & .850* \\
 \hline
\end{tabularx}%
\begin{tablenotes}
\item
*Correlation is significant at the 0.01 level (2-tailed).
\end{tablenotes}
\end{threeparttable}

\end{table}%

\begin{figure}[!tbp]
\caption{Comparison of the least squares and HDI rankings in 2015}
\begin{subfigure}{.495\textwidth}
\caption{Least squares ranking from remittances}
\centering
\includegraphics[scale=1]{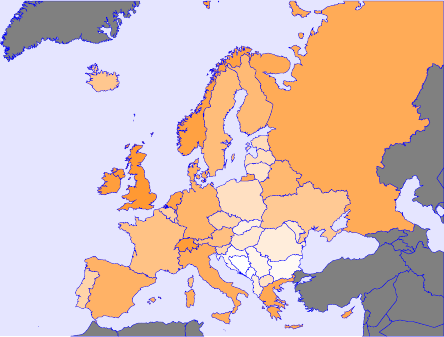}
%\begin{tikzpicture}[scale=0.135,background rectangle/.style={fill=blue!10},show background rectangle,inner frame xsep=-0pt,inner frame ysep=-0pt]
%%\input{Europe_political_colorable.tex}
%
%\tikzset{set country val/.style args={#1/#2}{#1={fill=orange!#2}}}
%\tikzset{set country val/.list={NO/74,
%CH/78,
%DE/58,
%DK/62,
%NL/72,
%IE/76,
%IS/44,
%SE/52,
%GB/80,
%LU/40,
%FR/48,
%BE/38,
%FI/54,
%AT/50,
%SI/30,
%IT/64,
%ES/60,
%CZ/32,
%GR/68,
%EE/28,
%CY/70,
%MT/36,
%PL/24,
%LT/20,
%SK/10,
%PT/34,
%HU/18,
%LV/26,
%HR/6,
%ME/2,
%RU/66,
%RO/14,
%BY/56,
%BG/4,
%RS/8,
%AL/12,
%BA/0,
%MK/22,
%UA/42,
%MD/16
%}}
%
%%\Europe[every country={draw=blue, line width=0.05mm, fill=black!50}]
%\end{tikzpicture}
\end{subfigure}
\begin{subfigure}{.495\textwidth}
\caption{HDI ranking}
\centering
\includegraphics[scale=1]{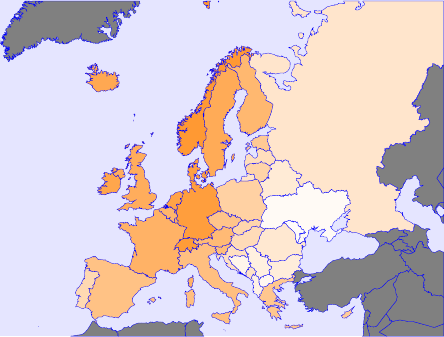}
%\begin{tikzpicture}[scale=0.135,background rectangle/.style={fill=blue!10},show background rectangle,inner frame xsep=-0pt,inner frame ysep=-0pt]
%
%%\input{Europe_political_colorable.tex}
%
%\tikzset{set country val/.style args={#1/#2}{#1={fill=orange!#2}}}
%\tikzset{set country val/.list={NO/80,
%CH/78,
%DE/76,
%DK/74,
%NL/72,
%IE/70,
%IS/68,
%SE/66,
%GB/64,
%LU/62,
%FR/60,
%BE/58,
%FI/56,
%AT/54,
%SI/52,
%IT/50,
%ES/48,
%CZ/46,
%GR/44,
%EE/42,
%CY/40,
%MT/38,
%PL/36,
%LT/34,
%SK/32,
%PT/30,
%HU/28,
%LV/26,
%HR/24,
%ME/22,
%RU/20,
%RO/18,
%BY/16,
%BG/14,
%RS/12,
%AL/10,
%BA/8,
%MK/6,
%UA/4,
%MD/2
%}}
%
%\Europe[every country={draw=blue, line width=0.05mm, fill=black!50}]
%\end{tikzpicture}
\end{subfigure}

\label{fig2}
\end{figure}

The HDI is also compared with our alternative quality of life ranking in Figure~\ref{fig2}. A darker colour indicates a higher rank. The results are similar for most countries. While the HDI places Norway at the top, the United Kingdom remains in a leading position. However, Belarus, Russia, and Ukraine get a significantly better rank with our methodology than shown by the HDI. Since a possible reason is that we have restricted our analysis to the European countries, and handle all others as one entity, this bias requires further investigation. 

\begin{table}[!ht]
\caption{Quality of life rankings by the least squares method with and without the CIS in 2015}
\label{Table4}
\centering
\begin{threeparttable}
\begin{tabularx}{\linewidth}{CcC CcC} \toprule %\hiderowcolors
Country  &  R &  R (CIS) & Country  &  R &  R (CIS) \\ \hline %\showrowcolors
    AL    & 35    & 33 \up & IS    & 19    & 18 \up \\
    AT    & 16    & 14 \up & 					IT   & 9    & {8 } \up \\
   	BA    & 41    & 42 \down &  LT    & 31    & 37 \down \\
    BE    & 22    & 21 \up & LU    & 21    & 19 \up \\
    BG    & 39    & 39 \const  & LV    & 28    & 34 \down \\
    BY    & 13    & 20 \down &MD    & 33    & 40 \down \\
   	CH    & 2     & {2 } \const  & ME    & 40    & 41 \down \\
    CY    & 6     & {6 } \const  & MK    & 30    & 30 \const \\
    CZ    & 25    & 25 \const  & MT    & 23    & 22 \up \\
    DE    & 12    & 11 \up & NL    & 5     & {4 } \up \\
	DK    & 10    & 10  \const & NO    & 4     & {5 } \down\\
    EE    & 27    & 27   \const  & PL    & 29    & 28 \up \\
    ES    & 11    & {9 } \up & PT    & 24    & 23 \up \\
	FI    & 14    & 12 \up & RO    & 34    &32  \up \\
    FR    & 17    & 15 \up & RS    & 37    & 36 \up \\
   	GR    & 7     & {7 } \const  & RU    & 8     & 17 \down \\
    GB 	  & 1     & {1 } \const & SE    & 15    & 13 \up \\
    HR    & 38    & 38  \const  & SI    & 26    & 24 \up \\
   	HU    & 32    & 31 \up  & SK    & 36    & 35 \up \\
    IE    & 3     & {3 } \const & UA    & 20    & 26 \down \\
    \hline
    Other & 18    & 16 \down & CIS   & ---     & 29 \, \, \\
    \hline
    \end{tabularx}
\begin{tablenotes}
\item
The column with the header ``R'' shows the original rank of the countries according to the least squares method, and the columns with the header ``R (CIS)'' corresponds to the case when the CIS (Commonwealth of Independent States) entity is introduced.
\end{tablenotes}
\end{threeparttable}
\end{table}%
 
Therefore, we have merged Armenia, Azerbaijan, Georgia, Kazakhstan, the Kyrgyz Republic, Tajikistan, Turkmenistan, and Uzbekistan into a so-called ``post-Soviet'' or CIS (Commonwealth of Independent States) entity, and repeated the computations. 

Table~\ref{Table4} compares the new rankings by the least squares method to the original one. Red arrow indicates a negative, green signs positive a change, while circle means no change. The ranking with the CIS entity can be found in Table~\ref{table5} in the Appendix.

According to the new ranking, Belarus, Latvia, Lithuania, Moldova, Russia, and Ukraine obtain a substantially worse position, but there are no visible improvements. The reason is that original ``Other'' entity is in the middle of the ranking, however, the CIS is low-ranked entity, thus the countries are rearranged.
According to this result, the use of the least squares method to country ranking may be sensitive to aggregation, and it requires properly disaggregated data. But this is not an inherent flaw of our proposal: \citet{Csato2019l} shows via an impossibility theorem that any reasonable ranking should depend on the level of aggregation.

\section{Conclusions} \label{Sec4}

This paper has considered remittances as quantification of preferences revealed by people working abroad, underlying a ranking of countries around the world. Nonetheless, the use of remittances can be criticised from various aspects because: (1) the data on migration in various destination countries are incomplete; (2) the incomes of migrants and the costs of living are proxied by per capita incomes in PPP terms; and (3) there is no way to capture remittances flowing through informal, unrecorded channels. %Nevertheless, it seems to be the best database available on international remittances.

These caveats somewhat limit the validity of our results. On the other hand, the proposed methodology has some advantages, illustrated by its independence of arbitrary parameter choices and favourable axiomatic properties. Its similarity to the Human Development Index indicates that we are able to capture at least some aspects of ``quality of life''. It can help evaluate a country's performance, even more finding a role model for emerging regions. For instance, among Post-Yugoslav states North Macedonia stands out, while in the Baltic region of the former Soviet Union, Estonia performs slightly better than Latvia and Lithuania.  Finally, our suggested ranking can serve as a baseline for other composite indices.

While our approach cannot immediately substitute other rankings, the suggested ranking may become an alternative to various composite indices as it is straightforward to calculate. Furthermore, it only requires data on remittances. Hopefully, the current research will contribute to a better understanding of economic and social development.

\section*{Acknowledgements}
\addcontentsline{toc}{section}{Acknowledgements}
\noindent
We are grateful to \emph{L\'aszl\'o Csat\'o}, \emph{Tam\'as Halm}, \emph{Istv\'an K\'onya}, \emph{Gy\"orgy Moln\'ar} and all participants of the \href{https://www.uni-corvinus.hu/index.php?id=afml_2018}{9th Annual Financial Market Liquidity Conference} for useful advice. Two anonymous reviewer provided valuable remarks and suggestions.

\noindent
This research was supported by the Higher Education Institutional Excellence Program of the Ministry of Human Capacities in the framework of the ``Financial and Public Services'' research project (20764-3/2018/FEKUTSTRAT) at Corvinus University of Budapest and the NKFIH grant K 128573.

\bibliographystyle{apalike}
\bibliography{All_references}

\section*{Appendix}\label{app}
\addcontentsline{toc}{section}{Appendix}

\begin{table}[!htbp]
  \centering
  \caption{The abbreviations of country names and the quality of life ranking by the least squares method with the CIS entity in 2015}
\rowcolors{1}{gray!20}{}
    \begin{tabularx}{0.6\linewidth}{lcc}
    \toprule \hiderowcolors
      Country & Abbreviation & R (CIS)  \\
\midrule \showrowcolors
     United Kingdom & GB    & 1 \\
    Switzerland & CH    & 2 \\
    Ireland & IE    & 3 \\
    Netherlands & NL    & 4 \\
    Norway & NO    & 5 \\
    Cyprus & CY    & 6 \\
    Greece & GR    & 7 \\
    Italy & IT    & 8 \\
    Spain & ES    & 9 \\
    Denmark & DK    & 10 \\
    Germany & DE    & 11 \\
    Finland & FI    & 12 \\
    Sweden & SE    & 13 \\
    Austria & AT    & 14 \\
    France & FR    & 15 \\
    Russian Federation & RU    & 17 \\
    Iceland & IS    & 18 \\
    Luxembourg & LU    & 19 \\
    Belarus & BY    & 20 \\
    Belgium & BE    & 21 \\
    Malta & MT    & 22 \\
    Portugal & PT    & 23 \\
    Slovenia & SI    & 24 \\
    Czech Republic & CZ    & 25 \\
    Ukraine & UA    & 26 \\
    Estonia & EE    & 27 \\
    Poland & PL    & 28 \\
    North Macedonia & MK    & 30 \\
    Hungary & HU    & 31 \\
    Romania & RO    & 32 \\
    Albania & AL    & 33 \\
    Latvia & LV    & 34 \\
    Slovak Republic & SK    & 35 \\
    Serbia & RS    & 36 \\
    Lithuania & LT    & 37 \\
    Croatia & HR    & 38 \\
    Bulgaria & BG    & 39 \\
    Moldova & MD    & 40 \\
    Montenegro & ME    & 41 \\
    Bosnia and Hercegovina & BA    & 42 \\ \hline
    Other &       & 16 \\
    CIS   &       & 29 \\  \hline
\end{tabularx}%
 \label{table5}
\end{table}

%\begin{table}[htbp!]
%  \centering
%  \caption{Quality of life rankings by the least squares method with the CIS in 2015}
%  %\rowcolors{1}{gray!20}{}
%  \begin{tabular}{cc} 
% \toprule %\hiderowcolors
%    
%   Country &  $\mathbf{q}(\mathbf{A})$ with CIS \\ 
%  \midrule %\showrowcolors 
%    GB    & 1 \\
%    CH    & 2 \\
%    IE    & 3 \\
%    NL    & 4 \\
%    NO    & 5 \\
%    CY    & 6 \\
%    GR    & 7 \\
%    IT    & 8 \\
%    ES    & 9 \\
%    DK    & 10 \\
%    DE    & 11 \\
%    FI    & 12 \\
%    SE    & 13 \\
%    AT    & 14 \\
%    FR    & 15 \\
%    RU    & 17 \\
%    IS    & 18 \\
%    LU    & 19 \\
%    BY    & 20 \\
%    BE    & 21 \\
%    MT    & 22 \\
%    PT    & 23 \\
%    SI    & 24 \\
%    CZ    & 25 \\
%    UA    & 26 \\
%    EE    & 27 \\
%    PL    & 28 \\
%    MK    & 30 \\
%    HU    & 31 \\
%    RO    & 32 \\
%    AL    & 33 \\
%    LV    & 34 \\
%    SK    & 35 \\
%    RS    & 36 \\
%    LT    & 37 \\
%    HR    & 38 \\
%    BG    & 39 \\
%    MD    & 40 \\
%    ME    & 41 \\
%    BA    & 42 \\ \hline
%    Other & 16 \\
%    CIS   & 29 \\
%    \hline
%    \end{tabular}
%  \label{tab_rank}
%\end{table}

\clearpage
\begin{table}[htbp]
  \centering
  \caption{Remittances by countries in 2015 (million USD)}
  \rowcolors{1}{gray!20}{}
\begin{tabularx}{\linewidth}{l  Rr rR}
\toprule \hiderowcolors
    Country & Outflow & Inflow & Net remittances & Net remittances/GDP \\ \midrule
    \showrowcolors
     AL   & 194.9 & 1047.0 & -852.1 & -7.5\% \\
     AT   & 3738.5 & 2814.0 & 924.5 & 0.2\% \\
     BY   & 835.1 & 695.7 & 139.4 & 0.2\% \\
     BE   & 5654.2 & 9934.0 & -4279.8 & -0.9\% \\
     BA   & 51.6  & 1771.8 & -1720.2 & -10.6\% \\
     BG   & 132.3 & 1443.1 & -1310.8 & -2.6\% \\
     HR   & 1145.1 & 2103.6 & -958.5 & -1.9\% \\
     CY   & 409.5 & 248.8 & 160.7 & 0.8\% \\
     CZ   & 2013.2 & 2692.9 & -679.7 & -0.4\% \\
     DK   & 1772.9 & 1246.8 & 526.1 & 0.2\% \\
     EE   & 176.4 & 445.5 & -269.2 & -1.2\% \\
     FI   & 921.1 & 806.3 & 114.8 & 0.0\% \\
     FR   & 20864.4 & 23347.1 & -2482.7 & -0.1\% \\
     DE   & 22967.3 & 15362.1 & 7605.2 & 0.2\% \\
     GR   & 1317.5 & 428.8 & 888.7 & 0.5\% \\
     HU   & 916.9 & 4021.0 & -3104.2 & -2.5\% \\
     IS   & 126.5 & 190.6 & -64.1 & -0.4\% \\
     IE   & 2026.0 & 601.2 & 1424.8 & 0.5\% \\
     IT   & 15487.4 & 9517.0 & 5970.4 & 0.3\% \\
     LV   & 278.3 & 1416.3 & -1138.1 & -4.2\% \\
     LT   & 208.9 & 1373.8 & -1165.0 & -2.8\% \\
     LU   & 1232.8 & 1613.0 & -380.2 & -0.7\% \\
     MT   & 81.2  & 167.8 & -86.6 & -0.8\% \\
     MD   & 296.2 & 1533.4 & -1237.2 & -16.0\% \\
     ME   & 66.4  & 381.2 & -314.8 & -7.8\% \\
     NL   & 5361.0 & 1364.9 & 3996.1 & 0.5\% \\
     MK   & 122.0 & 307.0 & -185.0 & -1.8\% \\
     NO   & 2235.4 & 609.8 & 1625.6 & 0.4\% \\
     PL   & 1489.3 & 6785.0 & -5295.7 & -1.1\% \\
     PT   & 2303.6 & 4367.7 & -2064.1 & -1.0\% \\
     RO   & 548.4 & 2932.5 & -2384.1 & -1.3\% \\
     RU   & 14647.4 & 6869.6 & 7777.8 & 0.6\% \\
     RS   & 1265.0 & 3370.7 & -2105.7 & -5.3\% \\
     SK   & 683.6 & 2137.6 & -1454.0 & -1.6\% \\
     SI   & 751.5 & 728.8 & 22.7  & 0.1\% \\
     ES   & 15850.9 & 10273.7 & 5577.2 & 0.5\% \\
     SE   & 3401.4 & 3268.7 & 132.7 & 0.0\% \\
     CH   & 8626.7 & 2234.9 & 6391.8 & 0.9\% \\
     UA   & 3795.6 & 5845.0 & -2049.4 & -2.3\% \\
     GB   & 25337.4 & 5003.4 & 20334.0 & 0.7\% \\ \midrule
    Sum   & 169333.4 & 141301.9 & 28031.5 &  ---\\
    Average & 4233.3 & 3532.5 & 700.8 & -1.7\% \\ \hline
    \end{tabularx}%

  \label{netrem}%
\end{table}%

\end{document}